# All-electric spin device operation using the Weyl semimetal, WTe$_2$, at room temperature

Kosuke Ohnishi[1], Motomi Aoki[1], Ryo Ohshima[1], Ei Shigematsu[1], Yuichiro Ando[1,2], Taishi Takenobu[3], and Masashi Shiraishi[1, #]

1. Department of Electronic Science and Engineering, Kyoto University, Kyoto, Kyoto 615-8510, Japan.

2. PRESTO, Japan Science and Technology Agency, Honcho, Kawaguchi, Saitama 332-0012, Japan.

3. Department of Applied Physics, Nagoya University, Nagoya 464-8603, Japan.

# Corresponding author: Masashi Shiraishi (shiraishi.masashi.4w@kyoto-u.ac.jp)

**Abstract**

Topological quantum materials (TQMs) possess abundant and attractive spin physics, and a Weyl semimetal is the representative material because of the generation of spin polarization that is available for spin devices due to fictitious Weyl monopoles at the edge of the Weyl node. Meanwhile, a Weyl semimetal allows the other but unexplored spin polarization due to local symmetry breaking. Here, we report all-electric spin device operation using a type-II Weyl semimetal, $WTe_2$, at room temperature. The polarization of spins propagating in the all-electric device is perpendicular to the $WTe_2$ plane, which is ascribed to local in-plane symmetry breaking in $WTe_2$, yielding the spin polarization creation of propagating charged carriers, namely, the spin-polarized state creation from the non-polarized state. Systematic control experiments unequivocally negate unexpected artifacts, such as the anomalous Hall effect, the anisotropic magnetoresistance etc. Creation of all-electric spin devices made of TQMs and their operation at room temperature can pave a new pathway for novel spin devices made of TQMs resilient to thermal fluctuation.

# 1. Introduction

Since the discovery of a new material phase, topological quantum materials (TQMs) [1-3], tremendous effort has been made to explore a wide variety of novel and abundant physics appearing in topological insulators (TIs), topological superconductors (TSCs), and Weyl semimetals for creating novel electric and spintronics devices by utilizing their fast carrier mobilities due to the linear band structures. Contrary to topologically trivial materials, TQMs exhibit great potential to show exotic and intriguing physical properties. TIs possess spin-polarized Dirac fermion bands resulting in the quantum spin Hall effect and persistent spin current that are resilient to defects [4], and TSCs can be a material platform for realizing fault-tolerant quantum computing utilizing Majorana fermions that can exist in TSC states [5]. A Weyl semimetal is a considerably new family of TQMs [6], where nondegenerate linear conduction and valence bands touch each other at the Weyl point by breaking either spatial or time reversal symmetries, yielding gapless Weyl fermions. In addition, a Weyl semimetal is known to be an ideal material platform of the appearance of the Adler-Bell-Jackiw (ABJ) anomaly of Weyl fermions [7], which is due to the breaking of chiral symmetry in massless Weyl fermions under quantum fluctuation. The ABJ anomaly was experimentally corroborated in TaAs, NbAs, TaP and $WTe_2$ [8-11] and is the fingerprint of the Weyl nature.

Importantly, fictitious magnetic monopoles can appear at each Weyl point, and the Weyl node connects both monopoles, which gives rise to in-plane spin polarization along the Weyl node at the surface of the Weyl semimetal, as theoretically proposed by Nielsen and Ninomiya [12,13]. Among Weyl semimetals, $T_d$-type $WTe_2$ is a new class, a type-II Weyl semimetal [14], where the Weyl points appear at the crossing of the oblique conduction and valence bands due to the broken inversion symmetry and nonsaturating giant positive magnetoresistance is a manifestation of the type-II Weyl character [15,16]. The quest for substantiating spin information extraction of the in-plane spin polarization in WTe$_2$ allowed successful detection of spin accumulation of the spin polarization along the *b*-axis ($S_y$, parallel to the WTe$_2$ plane) that is ascribed to the Weyl node by introducing an electrical (potentiometric) method. This work provides a significant breakthrough in a wide variety of electronics and spintronics using Weyl semimetals since all-electric spin generation and extraction are key for the creation of novel spin devices using TQMs. However, detection is limited at very low temperature (<15 K) [17] due to lattice expansion, which opens an energy gap at the Weyl point. Thus, electrical spin information propagation and extraction in Weyl semimetals for creating novel all-electric spin devices is almost unexplored in spite of expectation of quite fast spin propagation thanks to the Dirac-like linear band structure resulting in theoretically massless fermions, and consequently, the study is and largely lags behind that using TIs [18-21], where room temperature spin information extraction was realized.

Meanwhile, angle-resolved photoemission spectroscopy (ARPES) revealed other possible spin polarizations along the *c*-axis ($S_z$, perpendicular to the plane) in addition to those along the *b*-axis in WTe$_2$ [22]. This $S_z$ spin polarization in WTe$_2$ has not been utilized in all-electric spin devices, which hinders further progress of spintronic applications of Weyl semimetals and the creation of novel all-electric spin devices, especially because potentiometric detection of spin information in WTe$_2$ via spin accumulation has been strongly limited at very low temperatures, as mentioned above [17]. Here, in this work, we shed light on the spin polarization along the *c*-axis ($S_z$) of WTe$_2$, which is attributed to breaking of spatial inversion symmetry with respect to the *ac*-plane, i.e., symmetry breaking along the *b*-axis. The successful all-electric generation and extraction of the $S_z$ spin information is realized up to 300 K via spin accumulation, i.e., thermally robust. Additionally, the amplitude of the $S_z$ spin information signal due to its accumulated spins is almost one order of magnitude greater than that of $S_y$ (in-plane) due to the Weyl node. Furthermore, control experiments enable us to conclude that the $S_z$ spin polarization is definitely ascribed to in-plane inversion symmetry breaking of WTe$_2$, not to the interfacial Rashba effect arising at the interface of the ferromagnet (FM) and WTe$_2$, which indicates that the spin polarized state in the WTe$_2$ spin device is created from the non-polarized state, i.e., the successful spin creation. Both the Weyl node and $S_z$ spin polarization stem from in-plane inversion symmetry breaking, but our finding unequivocally reveals the resilience of the $S_z$ spin polarization in WTe$_2$.

## 2. Results and Discussion

$S_z$ spin polarization in WTe$_2$ is ascribed to the in-plane inversion symmetry breaking along the *b*-axis, resulting in substantial spin-orbit interaction (SOI) that enables an effective emergent magnetic field perpendicular to the WTe$_2$ plane. Contrary to previous work [17], we focus on $S_z$ spin polarization in WTe$_2$ and utilize an accomplished potentiometric method, where nonequilibrium $S_z$ spin accumulation in WTe$_2$ due to charge flow is detected, as in three-dimensional (3D) TIs [17-21]. Notably, the spin polarization direction at the surface of 3D-TIs is in-plane unlike that in WTe$_2$ because the SOI-induced effective emergent magnetic field in 3D-TIs is in-plane due to the Rashba field perpendicular to the surface plane of 3D-TIs. This enables exerting ferromagnetic electrodes with in-plane magnetization (such as single Co layer) to detect the surface spin polarization in 3D-TIs. Thus, unlike the study on spin detection in 3D-TIs, in the present study, a ferromagnetic detector electrode possessing perpendicular magnetic anisotropy (PMA) is equipped on a mechanically exfoliated WTe$_2$ thin film of 22 nm thick. The PMA electrodes are made of a [Co/Pt] multilayer, and Pt electrodes are also equipped as nonmagnetic electrodes (see Fig. 1(a) and **4. Experimental Section**). An external magnetic field to control the magnetization direction of the PMA electrodes is applied perpendicular to the WTe$_2$ plane.

The WTe$_2$ film for the all-electric spin device possesses a $T_d$ structure, which is supported by Raman spectroscopy (see Figs. 1(b) and 1(c)). Positive magnetoresistance is observed in the

WTe$_2$ of the spin device shown in Fig. 1(a) when an external magnetic field is applied along the *c*-axis, and a monotonic decrease in the resistance under cooling can be seen in the whole temperature range (see Figs. 1(d) and 1(e)). Thus, the WTe$_2$ in the device unequivocally possesses a Weyl semimetallic nature. The axis of electric current flow was determined by polarized Raman spectroscopy; for example, the current flows almost along the *a*-axis in the device shown in Fig. 1(a), as planned (see also Supporting Information No. 1).

Figure 2 shows central results of the study. The temperature evolution of the anomalous Hall effect (AHE) signal of a [Co/Pt]$_{10}$ PMA film and the measuring circuit to generate and extract $S_z$ spin information from the WTe$_2$ device are shown in Figs. 2(a) and 2(b), respectively (see also Supporting Information No. 2 for the AHE measurement). The PMA film exhibits ferromagnetic characteristics up to 300 K, where hysteresis due to the coercive force is noticeably observed. As mentioned above, accumulated $S_z$-polarized spin information beneath the PMA electrode is electrically generated by the charge flow along the *a*-axis and is extracted by controlling the magnetization direction of the PMA electrode. Hence, hysteresis in spin accumulation voltages can appear within the coercive force of the PMA electrode in the upward and downward sweeping of the external magnetic field perpendicular to the WTe$_2$ plane. In fact, salient hysteresis appears at 5 K and evolves up to 300 K (see Figs. 2(c) and 2(d)). The voltage hysteresis was successfully observed in the other devices using WTe$_2$ with different thicknesses (see Supporting Information No. 4 and No. 12). A concomitant voltage jump, for example, observed at ±(500-700) mT at 5 K, is attributed to the anisotropic magnetoresistance (AMR) of the PMA electrode (see also Supporting Information No. 3). The external magnetic fields, where the opening and closing of the hysteresis can be seen at each measuring temperature, are nicely consistent with the coercive forces. Additionally, the range of the external magnetic fields allowing the resistance hysteresis monotonically decreases as a function of temperature as does the hysteresis due to the AHE. Such temperature-dependent and noticeable hysteresis in the output spin voltages is rationalized by the mechanism discussed in the previous subsection. To note is possible superposition of artifact signals to the spin signals, for example, signals attributable to anisotropic magnetoresistance (AMR) and AHE of the PMA electrode, MR of the WTe$_2$ itself, a spurious signal due to the fringe field in the PMA electrode [23], and the interfacial effects as clarified in FM/semiconductor [24-27]. Thus, careful control experiments were carried out to negate contribution from the artifacts. Figure 2(e) shows a result of a similar measurement using only three nonmagnetic Pt electrodes, where no hysteresis like those observed from the device equipping a PMA electrode was observed. This result apparently eliminates the scenario that the MR of the WTe$_2$ gives rise to the spin signals. Further supporting evidence is the successful suppression of the parabolic offset voltages ascribed to the MR of the WTe$_2$ in the spin device possessing short gap length, where spin voltage hysteresis is sufficiently large (see Fig. S5 of Supporting Information No. 5). Figure 2(f) shows the results of the other control experiments, where a similar spin device possessing a Pt channel was used. The purpose of the introduction of the Pt channel is to decouple possible contribution of spin transport/accumulation that can take place in a Cu or Ag channel from the possible AHE-induced of the PMA electrode, i.e., the Pt channel device allows focusing only on the artifacts due to the AHE. The injection current was the same (1 mA) for both devices to realize the same current density for the cross-sectional area of the PMA electrode. Whilst weak voltage hysteresis can be seen in the Pt channel device, where the magnitude of the voltage hysteresis is about 2 μV, the voltage (2 μV) is one order of magnitude smaller than the spin voltage (~20 μV) observed in the WTe$_2$ spin device. Furthermore, it is corroborated that spin voltage hysteresis is successfully observed from the WTe$_2$ spin device even when the electric current is not injected into the PMA electrode in the "crossed" geometry and the AHE signal is negligibly small in the b-axis oriented WTe$_2$ spin device. These results unequivocally negate the possible contribution of the AHE of the

PMA electrodes to the hysteresis (see Supporting Information No. 8 and 9 for estimation of a magnitude of the AHE voltage using the Pt channel device). In addition, we also note that the results shown in Fig. 2(c) also bear out that the AMR does not generate resistance hysteresis, because the spin voltages at the positive and negative higher magnetic fields are not identical in the WTe$_2$ device. The remaining artifact is attributed to the fringe field from the FM electrode, which was systematically analyzed using Bi$_2$Se$_3$ equipping a Co electrode with in-plain magnetization [23]. That study shed light on significance of the fringe-field-induced Hall voltages in spin voltage hysteresis, where the polarities of the fringe fields were opposite each other at both sides of the Co electrode, resulting in hysteric behavior of spin voltages in sweeping of an external magnetic field. Meanwhile, the PMA electrode is used in our study, which allows the same polarities of the fringe fields at both sides of the PMA electrode. Furthermore, the result shown in Fig. 2(f) also unequivocally eliminates the possible fringe-field-induced Hall voltages in the hysteresis, since no hysteresis was observed in the PMA/Pt channel device. We also emphasize that the interface effects that can appear in FM/semiconductor spin devices under the same measuring setup [24-27] are also eliminated by the control experiments because the interface effects does not enable explaining the polarity change of the hysteresis by the reversal of the charge momentum, which is discussed in detail below (see also Fig. 3(a)). Hence, the results shown in Fig. 2(c) are compelling evidence for successful electrical $S_z$ spin information extraction and operation of an all-electric spin device made of WTe$_2$ up to 300 K, and the detected $S_z$ spin detection is quite robust and resilient to thermal fluctuation compared to $S_y$ spin polarization ascribed to the Weyl node [17].

In the following paragraphs, we discuss physics behind of the all-electric spin device operation. To bear out that the observed hysteresis stems from the successful $S_z$ spin information extraction from WTe$_2$, an additional measurement was conducted by inverting the charge current direction. The direction of the $S_z$ spin is determined by the polarity of the vector product of momentum $k$ and the symmetry breaking axis $n$ (|| the $b$-axis). Thus, when the direction of $k$ is inverted, the polarity of the spin voltages is reversed, yielding a reversal of the polarity of the spin voltage hysteresis. Furthermore, the polarization is proportional to the momentum $k$, resulting in a linear dependence on the magnitudes of injected electric current (i.e., an applied electric field) within a low excitation regime. Figure 3(a) shows the comparison of the dependences of the hysteresis polarity of the spin information extraction signals on the electric current directions. The polarity of the hysteresis is nicely reversed when the current direction is switched from positive to negative, which is the manifestation of spin polarization switching, as in TIs [18,20,21], but due to the absence of local inversion symmetry in WTe$_2$. Figure 3(b) shows the electric current dependence of spin voltage amplitudes, where the amplitude of the spin voltages is defined as the difference of the voltages at 0 mT in the downward and the upward sweep of the external magnetic field (the spin voltage under the upward sweeping is set as the standard voltage. See also Supporting Information No. 6 for the whole dataset of the measurements). To note is that only the k-vector direction of the flowing charge is reversed in the measurement without changing the measuring circuit as shown in the inset of the figure. Salient linear dependence can be seen within ±1 mA, as expected, and the maximum amplitude of the spin voltage is approximately 30 µV, which is almost one order of magnitude greater than that at the same temperature in a previous study using the in-plane $S_y$ spin polarization due to the Weyl node (4 µV, 5 K) [17]. The results shown in Fig. 3 are compelling evidence that the physics behind of all-electric spin device is creation of the $S_z$ spin polarization by local symmetry breaking and charge flow, i.e., the spin polarized state is created from the non-polarized state (spin amplification).

The possible underlying physics of the spin polarization perpendicular to the WTe$_2$ plane was proposed as follows: (i) local symmetry breaking, (ii) formation of hybridized electronic states between WTe$_2$ and the adjacent NiFe, and (iii) generation of spin-transfer torque in the NiFe by charge flow and carrier scattering at the interface of the WTe$_2$ and NiFe, in the previous study using ST-FMR [28]. Our potentiometric approach unequivocally negates the generation of spin torque in FM. To clarify the underlying physics in more detail, we prepared another type of a spin device by inserting a Cu layer (10 nm thick), a weak SOI material, between WTe$_2$ and PMA to control the strength of the possible Rashba field. Figures 4(a) and 4(b) show the results of spin voltage detection using PMA and PMA/Cu electrodes fabricated in the same WTe$_2$ device, respectively, where discernible and similar hysteresis in the spin voltages with almost the same signal amplitudes can be seen. The hysteresis observed in PMA/Cu/WTe$_2$ was not rectangular enough compared to that observed in PMA/WTe$_2$, which is attributable to suppression of the magnetic property of the PMA electrode due to the surface roughness of the inserted thin Cu film. However, the successful observation of the hysteresis in the PMA/WTe$_2$ and PMA/Cu/WTe$_2$ devices underscores the validity of the physics behind the generation of the $S_z$ spin polarization to be local symmetry breaking, not an interfacial Rashba field at FM/WTe$_2$.

## 3. Conclusion

Local symmetry breaking in a wide variety of material systems has allowed the creation of an emergent effective magnetic field in GaAs [29], Bi/Ag [30], bilayer graphene [31,32], van der Waals heterostructures [33,34], and Si/SiO$_2$ [35]. The results presented here enable Weyl semimetals to join a new family of intriguing material systems allowing spin creation, propagation and extraction like those materials systems described above. Furthermore, creation of an all-electric spin device made of a TQM allowing quite fast carrier motion due to the Dirac-like linear bands and its operation at room temperature can pave a new pathway for novel and robust spin devices resilient to thermal fluctuation.

## 4. Experimental Section

A commercially available $T_d$-type $WTe_2$ crystal (2D semiconductors) was used for fabrication of all-electric spin devices. $WTe_2$ thin flakes were mechanically exfoliated and transferred onto a $SiO_2$(100 nm)/Si substrate. $Ar^+$ milling was carried out to remove a possible oxidized layer on the $WTe_2$ surface after exfoliation. The PMA electrodes made of $[Pt/Co]_{10}$ ("10" denotes a stacking number) and nonmagnetic Pt electrodes were deposited on the $WTe_2$ flakes on the substrate by a sputtering method, where the thicknesses of Co and Pt of the PMA film were 1 and 4.1 nm, respectively. The thickness of the nonmagnetic Pt electrodes was set to 50 nm. The growth rates of Co and Pt were 1.5 nm/min and 0.56 nm/min, respectively, and the thickness of electrodes was estimated from these growth rates. The thickness of the $WTe_2$ thin film was measured by using atomic force microscopy (AFM), and the thickness of the device shown in Fig. 1(a) was measured to be approximately 50 nm. Device fabrication was implemented by using electron beam lithography.

The Raman spectroscopy measurements were implemented by using the Raman division system (HORIBA HR-800). The wavelength of the excitation laser was 632.8 nm, and the parallel-polarized configuration was used for the polarized Raman spectroscopy measurements. The magnetoresistance measurements were carried out by using the conventional two probe method. The outer two Pt electrodes were used as the source and the drain electrodes. The magnetoresistance was measured at 5 K by applying a dc electric current of 1 mA, and the resistance shown in Fig. 1(e) was determined from the $I$-$V$ curves at each temperature. For the spin voltage and magnetoresistance measurements, a physical property measurement system (Quantum Design) was used.

**Figures and figure captions**

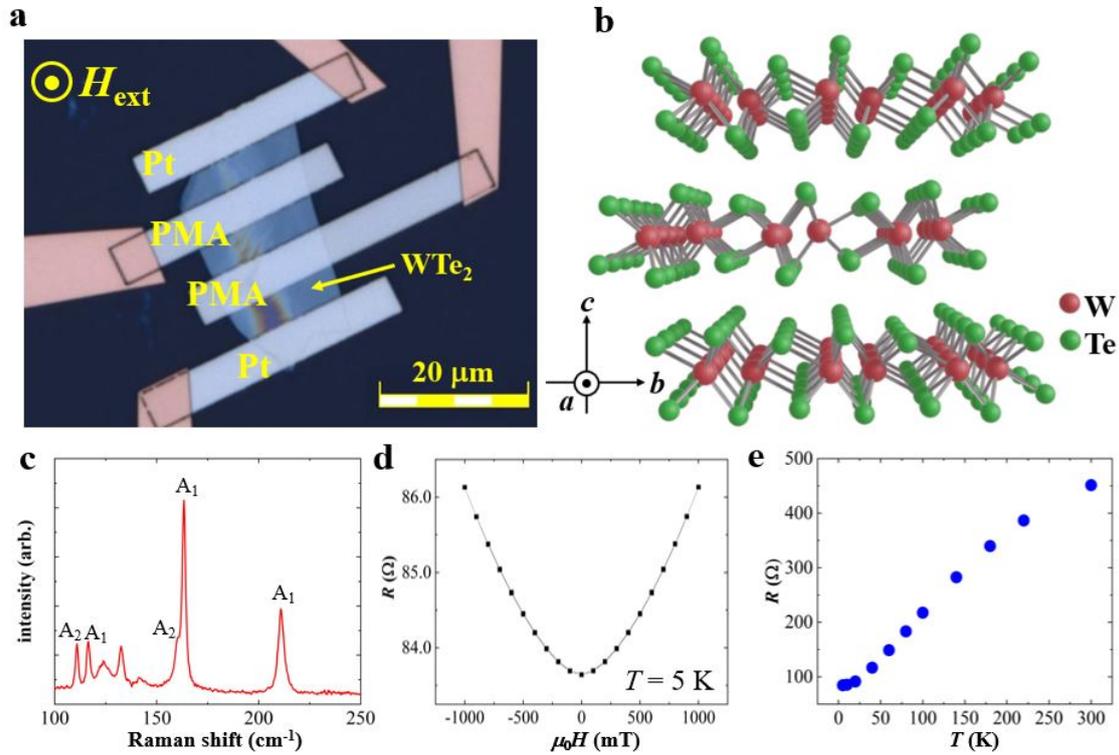

**Figure 1 (a)** Optical microscopic image of an all-electric spin device made of $WTe_2$. The inner two electrodes consist of ferromagnets with PMA, and the outer two electrodes consist of nonmagnetic Pt. The PMA is realized by a [Pt/Co] multilayer, where the thicknesses of Pt and Co are 4.1 and 1.0 nm, respectively, and the stacking number is 10. An external magnetic field is applied perpendicular to the $WTe_2$ plane. **(b)** Schematic image of $T_d$-type $WTe_2$. **(c)** Raman spectrum from the $WTe_2$. The identified Raman modes are denoted in the figure. **(d)** Magnetoresistance effect of the $WTe_2$. Large positive magnetoresistance under the application of an external magnetic field along the $c$-axis is a manifestation of a type-II Weyl semimetallic nature. **(e)** Temperature dependence of the two-terminal resistance of $WTe_2$. A monotonic decrease in the resistance supports the type-II Weyl semimetallic nature of $WTe_2$. The results shown in **(c)-(e)** were obtained from the spin device shown in **(a)**.

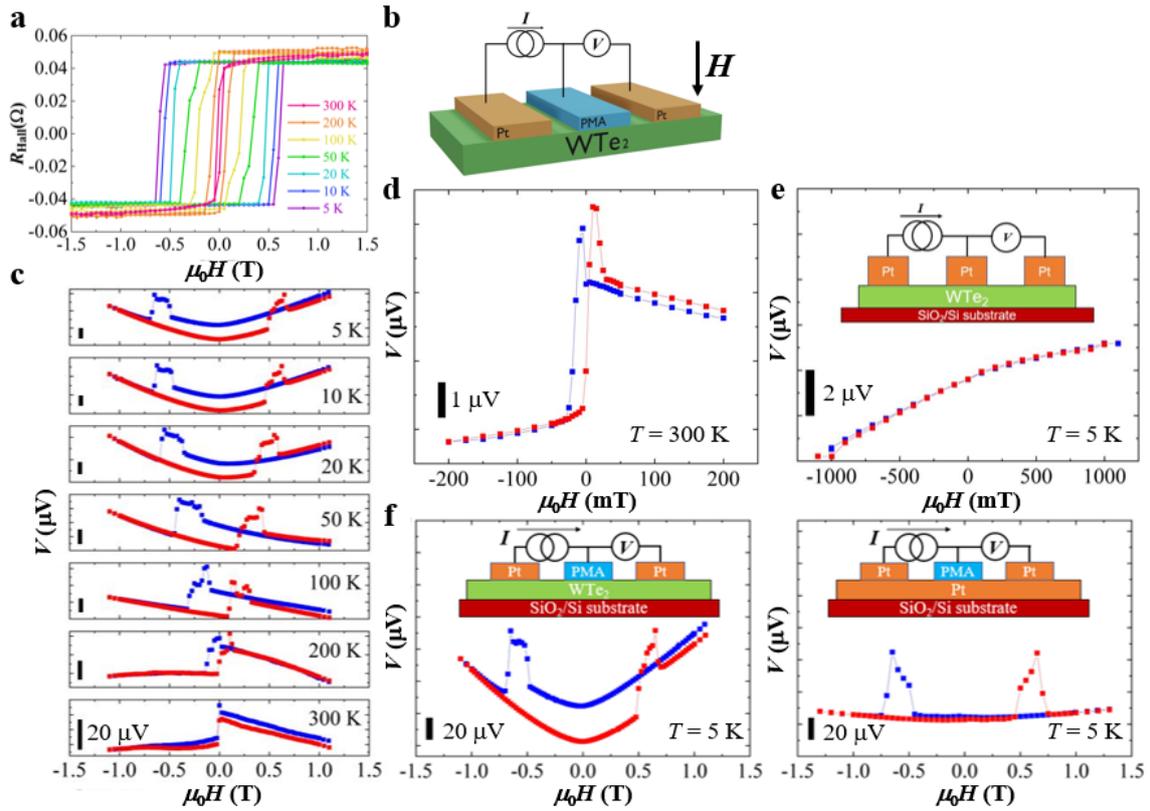

**Figure 2 (a)** Temperature evolution of the hysteresis due to the anomalous Hall effect (AHE) of the PMA electrode. The coercive force and saturation magnetization can be seen at 300 K. **(b)** Schematic image of the measuring setup of spin extraction from a $WTe_2$ spin device. Spin accumulation beneath the PMA is electrically created and potentiometrically detected. **(c)** Temperature evolution of spin voltage hysteresis observed from the $WTe_2$ spin device. Red (blue) closed squares are experimental data in the upward (downward) sweeping. While superposition of the AMR can be seen at each temperature, salient hysteresis attributed to successful spin detection by the PMA can be observed. The range of the hysteresis is consistent with the range of the AHE of the PMA. Black solid bars show the scale of 20 μV, and the injected electric current was set to be 1 mA. **(d)** Magnified view of the spin voltage hysteresis observed at 300 K. **(e)** A result of a control experiment using only Pt electrodes. The amplitude of the injected electric current was set to be 50 μV, and the measuring temperature was 5 K. Neither hysteresis nor AMR can be seen, which is compelling evidence that the hysteresis of the spin voltages is ascribed to the successful creation and detection of the $S_z$ spin polarization via spin accumulation. The inset shows the measuring setup. (f) Comparison of results for spin extraction measurements using the WTe$_2$ and Pt channels, respectively. The injection current was set to be 1 mA for both devices and the measuring temperature was 5 K.

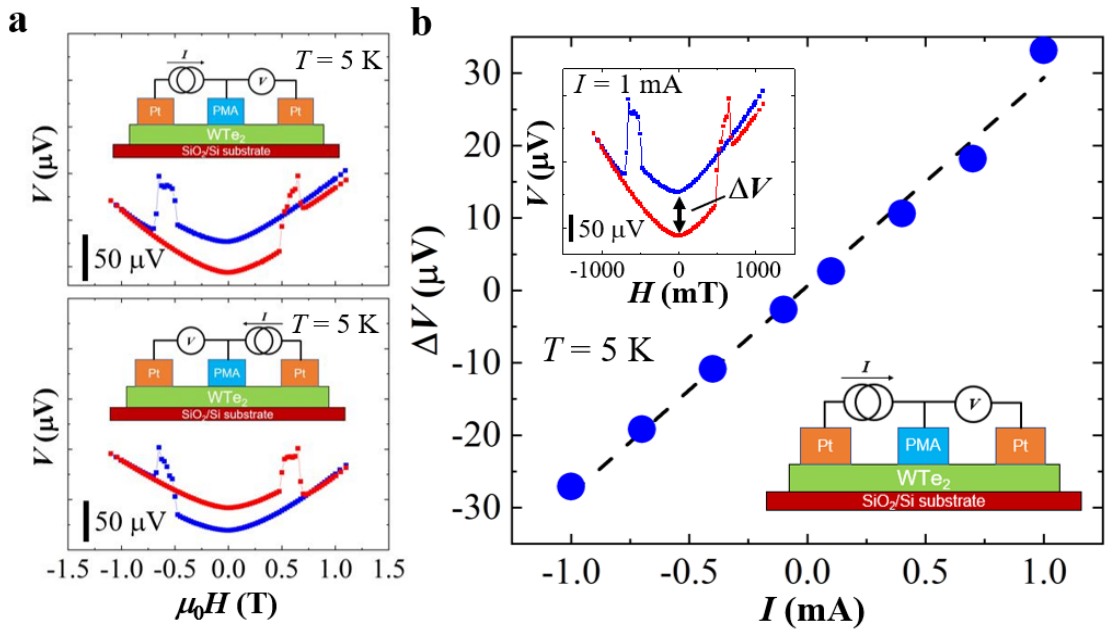

**Figure 3 (a)** Reversal of polarity of spin voltage hysteresis by switching the electric current direction. The red (blue) closed circles show the results under upward (downward) sweeping of the external magnetic field. The electric current directions (the sign of the *k*-vector of the spin carrier) are opposite in the setup shown in the upper and lower panels, and concomitantly, the polarities of the hysteresis are opposite. The inset shows the measuring setup, and the black scale bar shows 50 μV. The measuring temperature was 5 K and the injected electric current was 1 mA. **(b)** Injection current amplitude dependence of the output spin voltages (Δ*V*). Amplitudes of the output spin voltages are defined as the difference of the voltages at 0 mT on the basis of the voltage in the upward sweeping. The closed circles and the solid line are experimental data and the linear fitting, respectively. Noticeable linear dependence is observed within ±1 mA.

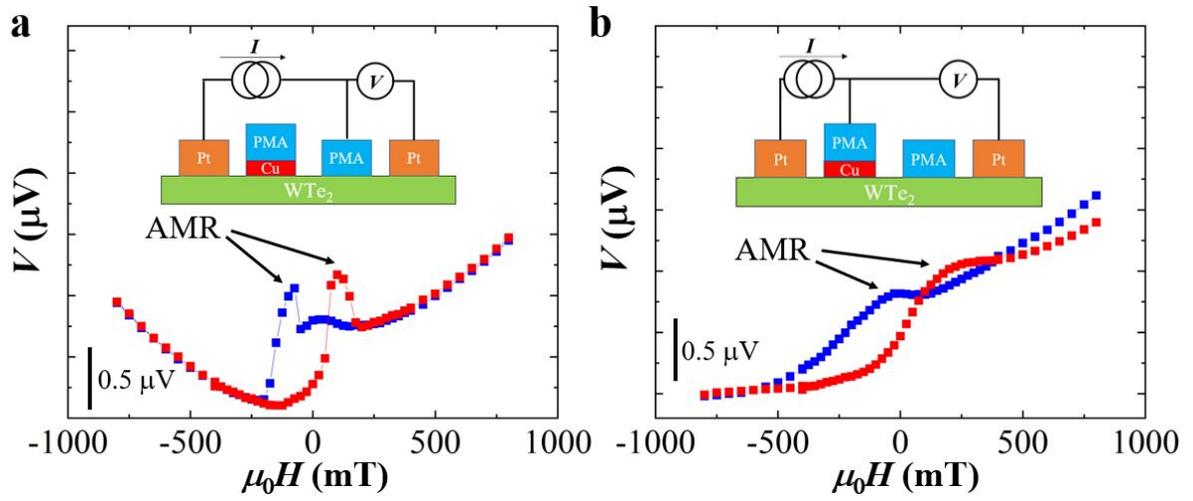

**Figure 4** Results of the spin voltage measurements using **(a)** PMAWTe$_2$ and **(b)** PMA/Cu/WTe$_2$. Hysteresis in the spin voltages can be observed in both setups. AMR is also seen in both measurements, whereas it is weak in PMA/Cu/WTe$_2$.

## Associated Content

**Supporting information**

Supporting information describes determination of the crystal axis with polarized Raman spectroscopy, Hall device fabrication for measuring the anomalous Hall effect (AHE) of the [Pt/Co] perpendicular magnetic anisotropy film, the effect of anisotropic magnetoresistance of PMA electrodes, reproducibility of the spin voltage measurement, the origin of the offset signals in the spin voltage measurement, and the whole data set of the electric current dependence of the spin voltages. The following files are available free of charge via the Internet (http://XXX).

**Corresponding Author**

Masashi Shiraishi, Department of Electronic Science and Engineering, Kyoto University, 6158510 Kyoto, Japan (E-mail: shiraishi.masashi.4w@kyoto-u.ac.jp)

**Author contributions**

M.S., M.A., K.O., and T.T. conceived the experiments. K.O. fabricated samples and collected data. K.O., M.A., R.O., E.S., Y.A. and M.S. analyzed the results. M.S. and K.O. wrote the manuscript. All authors discussed the results.

**Note**

The authors declare no competing interests.

**Acknowledgments**

This research is supported in part by the Japan Society for the Promotion of Science (JSPS) Grant-in-Aid for Scientific Research (S) (No. 16H06330), Grant-in-Aid for Scientific Research (A) (No. 21H04561), and JSPS Grant-in-Aid for Challenging Research (Pioneering) (No. 20K20443).